\documentclass[12pt]{article}
\usepackage{amsmath}
\usepackage{amssymb}
\usepackage{mathtext}
\usepackage{graphicx}

\begin{document}
\begin{centering}
\section*{CANONICAL GIBBS DISTRIBUTION AND\\ THERMODYNAMICS OF
MECHANICAL\\
SYSTEMS WITH A FINITE NUMBER \\OF DEGREES OF FREEDOM\footnotemark}
\footnotetext{REGULAR AND CHAOTIC DYNAMICS, V. 4, No. 2, 1999

{\it Received Jule 28, 1999

AMS MSC 82C22, 70F07}}
\end{centering}
\begin{centering}
V.\,V.\,KOZLOV\\
Faculty of Mechanics and Mathematics\\
Department of Theoretical Mechanics\\
Moscow State University\\
Vorob'ievy gory, 119899 Moscow, Russia\\
E-mail: vvkozlov@uni.udm.ru\\
\end{centering}\bigskip

\begin{abstract}
Traditional derivation of Gibbs canonical  distribution and the justification of thermodynamics are
based on the assumption concerning an isoenergetic ergodicity of a system of $n $ weakly interacting
identical subsystems and passage to the limit $n \to \infty $.
In the presented work we develop another approach to these problems assuming that $n$ is
fixed and $n \ge 2 $. The ergodic hypothesis (which frequently is not valid due to known results of
the KAM-theory) is substituted by a weaker assumption that the perturbed system does not have
 additional first integrals independent of the energy integral. The proof of nonintegrability of
perturbed Hamiltonian systems is based on  the Poincar\'e method.
Moreover, we use the natural
Gibbs assumption concerning a thermodynamic equilibrium of subsystems at vanishing interaction.
The general results are applied to the system of the weakly connected pendula.
The averaging with respect to the
Gibbs measure allows to pass from usual dynamics of mechanical systems to the classical
thermodynamic model.
\end{abstract}

\section{Introduction}

The classical approach to the justification of thermodynamics is based on the use of
{\it Gibbs canonical  distribution}
$$
\rho (x, y) = \frac{e ^ {-\beta H} }{ \int e ^ {-\beta H}\, dx\,dy }\, .
\eqno {(1.1)}
$$
Here $x $ and $y $ are canonical coordinates and momenta respectively, $H(x, y)$
is the Hamiltonian function  of a mechanical system,
$ \beta = \text {const} $. The function $ \rho $ is treated as a stationary
 density of probability distribution.
More precisely, we suppose that the system in the given state is a random event, and the
probability of detection of the system in area $D $ of the phase space is equal  to
$$
\int\limits_{D} \rho (x, y) \, dx\,dy\,.
$$

The parameter $ \beta $ is determined from the equality
$$
E = \int H \rho \, dx\,dy\,,
\eqno {(1.2)}
$$
\vspace*{-4mm}\\
where $E $ is average energy of the system. Usually this parameter is supposed to be
$ \beta = \frac{1}{kT} $, where~$T $ is the absolute temperature, $k $ is the Boltzmann
 constant \cite{1}--\cite{4}.
We assume that the integrals~\thetag{1.1} and~\thetag {1.2},
extended to the whole phase space, are
converging.

The justification of Gibbs distribution is one of basic problems of statistical mechanics. For this
purpose one usually considers $N $ {\it indistinguishable} systems with
the Hamiltonians
\vspace*{-1mm}
$$
H (x ^ {(\alpha)}, y ^ {(\alpha)})\,, \qquad \alpha = 1, \ldots, N\,,
\eqno {(1.3)}
$$
\vspace*{-6mm}\\
where $x ^ {(\alpha)} $, $y ^ {(\alpha)} $ ($ 1 \le i \le n $) are canonical coordinates of a system
with number $ \alpha $.
Let us emphasize that $n $ and $H $ do not depend on $ \alpha $.

Let us introduce the unified phase space of dimension $2nN $ with canonical variables
$$
( X, Y) = (x ^ {(1)}, y ^ {(1)}, \ldots, x ^ {(N)}, y ^ {(N)})
$$
\vspace*{-6mm}\\
and  the Hamiltonian
\vspace*{-2mm}
$$
\mathcal H _ {\varepsilon} (X, Y) = \sum _ {\alpha = 1} ^ {N} H (x ^ {(\alpha)}, y ^ {(\alpha)}) +
\varepsilon H_1 (X, Y, \varepsilon)\,, \eqno {(1.4)}
$$
where $ \varepsilon $ is a small parameter, which then will approach to zero. The system
with the Hamiltonian~\thetag{1.4} describes the dynamics of $N $
weakly interacting
subsystems with the Hamiltonians~\thetag{1.3}, and the function
$ \varepsilon H_1 $  denotes  the interaction energy  of subsystems.

The basic idea of derivation of Gibbs distribution is the assumption
concerning the {\it ergodicity} of the
system with the Hamiltonian~\thetag{1.4} at $ \varepsilon \ne 0 $ on a level surface $ \mathcal H _
{\varepsilon} = N E $~\cite{3}. Let $f $ be an integrable function on a phase space of a system
with $n $ degrees of freedom. According to an indistinguishability principle, in statistical
mechanics one considers
only such functions of $X $ and $Y $, which do not vary under
permutations of groups of variables $x^{(\alpha)} $, $y^{(\alpha)} $. Let us assume that
$$
F (X, Y) = \frac {1} {N} \sum _ {\alpha} f (x ^ {(\alpha)}, y ^ {(\alpha)})\,.
\eqno {(1.5)}
$$
Such functions are frequently called {\it summators}.
Using ergodic hypothesis, we obtain for theaverage
with respect to time \thetag {1.5} in the limit
as $ \varepsilon \to 0 $ the formula
$$
\langle F\rangle  \, = \int f (x, y) \rho_N (x, y) \, dx\, dy\,.
$$
The explicit expression for the density $ \rho_N $ can be found in \cite {3}. With some additional
suppositions \cite {3} we can prove that $ \rho_n \to \rho $ as $N \to \infty $ and for
the average of function $f $ along solutions of the system with the Hamiltonian \thetag {1.3} the formula
\vspace*{-3mm}
$$
\langle f\rangle  \, = \int f \rho \, dx\, dy
$$
\vspace*{-6mm}\\
is valid.

On this way one has a series of fundamental difficulties.
The main difficulty is the justification of the ergodic hypothesis.
Only recently the ergodicity of some simplified models has been established~\cite {5}. Since the Gibbs distribution \thetag
{1.1} does not depend on the form of an interaction energy~$H_1 $,
F.\,A.\,Berezin \cite{3} stated the idea of weakening of the ergodic
hypothesis: it is enough to require ergodicity of the
system with the Hamiltonian \thetag {1.4} for
$ \varepsilon \ne 0 $ for  set of
functions~$H_1$ everywhere dense.
However, if the  interaction energy~$H_1 $ has
no singularities and surfaces of the
energy level  $ \mathcal H_{\varepsilon} =N E $ are compact,
the ergodic hypothesis (even in the weakened variant)
isrefuted by the KAM-theory \cite {6}.

EXAMPLE 1. Let us consider a system with the Hamiltonian, following \cite {3}
$$
\mathcal H = \frac{1}{2} \sum (x _ {\alpha} ^2 + \omega^2 y _ {\alpha} ^2) + \varepsilon V_4 (x)\,,
\eqno {(1.6)}
$$
where $ \omega = \text {const} \ne 0 $, $V_4 \ge 0 $ is a polynomial of the 4-th degree of coordinates
$x_1, \ldots, x_N $, $ \varepsilon > 0 $. Such systems play an essential role in the theory of
heat capacities of rigid bodies \cite {2}.
The coefficients of non-negative polynomials of the 4-th degree form some
set $A $ in a
finite-dimensional space of coefficients of all polynomials of the 4-th
degree from $N $ variables.
F.\,A.\,Berezin~\cite {3} has raised the following question: Is it correct, that for almost all points $A$ the
system with the Hamilto\-nian~\thetag{1.6} is ergodic?
The answer  is negative. Indeed, let us consider polynomials of the form
\vspace*{-1mm}
$$
V_4 = \sum _ {\alpha} a _ {\alpha} x _ {\alpha} ^4 + W_4 (x)\,, \quad
a _ {\alpha} > 0\,,
$$
\vspace*{-4mm}\\
and let the coefficients of the polynomial $W_4 $ be small. If $W_4 = 0 $, the
 system with the Hamiltonian~\thetag {1.6} is integrated by a separation of variables. If we change to
 action-angle variables $I _ {\alpha} $, $ \varphi _ {\alpha} $ in a system with one degree of freedom
\vspace*{-2mm}
$$
h _ {\alpha} = \frac{(x^2 + \omega^2 y^2) }{2} + \varepsilon a _ {\alpha} x^4\,,
$$
\vspace*{-3mm}\\[-1mm]
then
\vspace*{-2mm}
$$
\mu _ {\alpha} = \frac{\partial h _ {\alpha} }{ \partial I _ {\alpha}} > 0\,, \qquad \lambda _ {\alpha} = \frac{\partial^2 h _
{\alpha} }{ \partial I _ {\alpha} ^2} > 0\,.
\eqno {(1.7)}
$$
In the last inequality we essentially use the supposition, that $ \varepsilon a _ {\alpha} > 0 $.

Let $H_0 = \sum h _ {\alpha} (I _ {\alpha}) $. If
\vspace*{-2mm}
$$
\Vmatrix
\frac {\partial^2 H_0} {\partial I^2} & \frac {\partial H_0} {\partial I} \\ \frac {\partial H_0}
{\partial I} & 0  \endVmatrix
\ne 0\,,
\eqno {(1.8)}
$$
then, according to the KAM-theory \cite {6}, the system with the Hamiltonian \thetag {1.6} is not
ergodic on each positive  energy level
for small $W_4 $ (when
coefficients of~$W_4 $ are
located in some small neighborhood of zero):
the most part of this surface is foliated on
$N $-dimensional invariant tori with conditional-periodic trajectories. In our case
the determinant
\thetag {1.8} is equal  to
$$
- \lambda_1 \ldots \lambda_n \bigg(\frac{a_1^2 }{ \lambda_1} + \ldots +
\frac{a_n^2 }{ \lambda_n}\bigg)\,,
$$
what does not equal zero on account of \thetag {1.7}.

It is still necessary to add that according to
the above mentioned  approach to the derivation of the Gibbs
distribution (due to Darwin and Fouler (see~\cite {4})) the expression for the
density of probability distribution \thetag {1.1} does
not depend on the interaction of subsystems. While such an
interaction is always present.

In this work we develop another approach to the justification of the formula for probability
density due to classical works  by Gibbs~\cite{1}. This approach will be based on the theory of
integrability of Hamiltonian systems \cite {7}.

\section{The main result}
The statistical approach to the theory of dynamical
systems assumes a refusal of consideration of
separate trajectories.
Instead, in the phase space of a system
$$
\dot x = v (x)
\eqno {(2.1)}
$$
we introduce the density of probability distribution $ \rho (x, t) > 0 $,
 explicitly depending on time in general case.
Let $D $ be any measurable area of the phase space,
$g^t $ be a phase flow of the system
\thetag {2.1}. This flow can be  presented as a stationary flow of a fluid.
Since the area $g^t (D) $ consists of the same moving phase points, it is natural to assume that
the probability of detection of the system in the area $g^t (D) $ does not depend on $t$. Hence, this
probability
$$
\int\limits _ {g^t (D)} \rho (x, t) \, dx
$$
will be an integral invariant of the system \thetag {2.1}. But then the  density of probability
distribution satisfies  the Liouville equation:
$$
\frac{\partial \rho }{ \partial t} + \text {div} (\rho v) = 0\,.
$$

If $ \text {div}\, v = 0 $, then $ \rho $ is a first integral of the system \thetag {2.1}.
This condition, obviously, holds for the Hamiltonian systems of differential equations.
It is natural to consider stationary distributions
for autonomous systems, when $ \rho $ does not
depend on $t $  explicitly. The detailed discussion of these problems can be found in \cite {2, 4}.

Locally, in a small neighbourhood of a nonsingular point the dynamical system \thetag {2.1} possesses a
complete set
of independent first integrals
(in  amount of $m-1 $, where $m $ is a dimension of the phase space).
However, in a typical situation they can not be extended to single-valued integrals defined
in the whole
phase space.
The density of probability distribution  is a single-valued function on the phase
space by  definition. Generally Hamiltonian systems with compact level surfaces do not admit integrals
independent of the Hamiltonian $H $, though they are not ergodic (on energy levels) \cite{7}.
This argument allows  to state at once in some important cases, that
$$
\rho = f (H)\,,
\eqno {(2.2)}
$$
and to reduce the problem to determination of the form of function $f $.

However, the opposite point of view to the possibility of existence of
additional integrals  is widespread in physical literature (see, for
example, \cite{4}, where the approach to justification
of Gibbs distribution based on the formula \thetag {2.2}
is called ``speculative").

To show the possibilities of a new approach, let us consider a Hamiltonian
 system with $n $ degrees of freedom. The Hamiltonian of this system
has the form \thetag {1.4}:
$$
H = H_0 + \varepsilon H_1 + o (\varepsilon)\,, \quad
H_0 = \sum h_i (x_i, y_i)\,, \quad H_1 = H_1 (x_1, \ldots, x_n, y_1, \ldots, y_n)\,. \eqno {(2.3)}
$$
At $ \varepsilon = 0 $ the system splits into $n $ independent subsystems with one degree of
freedom. A system of connected pendula can be considered as an example.

Let us assume that in each one-dimensional system with a Hamiltonian $h_i $
($ 1 \le i \le n $) we can introduce the canonical action-angle variables $I_i $, $ \varphi _i \, (\text
{mod}\, 2\pi) $~\cite {6}.
For example, for a system with the Hamiltonian $h _ {\alpha} $ from paragraph~1 such variables
are defined in all phase space. In case of
a pendulum the cylindrical phase space is divided by
separatrices in three areas, in each area it
is possible to introduce  action-angle variables. In
these variables
$$
h_i = h_i (I_i)\,, \quad 1 \le i \le n\,.
\eqno {(2.4)}
$$

Let us assume that the functions \thetag {2.4} are continuous and monotonically increasing. The
naturality of this supposition follows from the definition of an action variable as a normalized area
on a phase plane contained inside a phase curve $h_i = \text {const} $. Then the variable $I_i $ will
be a single-valued function of energy $h_i $.

The above mentioned assumptions are not fulfilled in the case, when the potential energy of a system has
some local minimums. However, these conditions have a technical character, and probably they
can be essentially  weakened.

Thus, in action-angle variables the  Hamiltonian \thetag {2.3} has a form
$$
H = \sum _ {i=1} ^ {n} h_i (I_i) + \varepsilon H_1 (I_1, \ldots, I_n, \varphi_1, \ldots, \varphi_n) + o
(\varepsilon)\,.
\eqno {(2.5)}
$$
It is $ 2 \pi $-periodic with respect to each angular variable $ \varphi_1, \ldots, \varphi_n $.

The  density of probability distribution $ \rho $ for a system with the Hamiltonian \thetag {2.3} is
a single-valued positive function of $x $, $y $, depending on parameter $ \varepsilon $. Let us assume
that $ \rho $ is a function from class $C^2 $ (it possesses the second continuous derivative on a set of
$2n+1 $ variables $x $, $y $, $ \varepsilon $). At small values
of~$ \varepsilon $ this function can be written as
\vspace*{-3mm}
$$
\rho = \rho_0 (x, y) + \varepsilon \rho_1 (x, y) + o (\varepsilon)\,,
\eqno {(2.6)}
$$
\vspace*{-3mm}\\[-2mm]
where $ \rho_0 $ and $ \rho_1 $ are functions from classes $C^2 $ and $C^1 $
respectively.

Now let $ \varepsilon $ tend to zero. Then $ \rho_0 $ will be a density  of probability distribution for a
system with the Hamiltonian
$$
H_0 = h_1 (x_1, y_1) + \ldots + h_n (x_n, y_n)\,.
$$

This system with $n$~degrees of freedom splits into $n$~{\it independent}
systems with one degree of freedom. The main independence property denotes that
the motion of each of these subsystems is uniquely determinated by any of its
initial states.

If we stick to the idea of statistical description of dynamical systems,
it is necessary to introduce  stationary densities of probability distribution
$$
p_1(x_1, y_1), \ldots , p_n(x_n, y_n)
$$
for each of one-dimensional systems. Taking into consideration the independence
properties and using the probability product theorem, we obtain
$$
\rho_0 = p_1 \ldots p_n\,.
\eqno{(2.7)}
$$

This equality is often called {\it Gibbs hypothesis on
thermodynamic equilibrium}~\cite{2}.

{\bf Remark 1.} {\it One should not think that any separation of variables results in
independent systems with one degree of freedom.  Let us show this fact by a simple
example}~:
$$
2 H_0 = y_2^2 + x_2^2 \left[\frac{(y_1^2 + x_1^2)}{2} \right]^2.
$$
{\it The variables~$x_2$, $y_2$ perform simple harmonic oscillations, the frequency
of which is equal to the energy of oscillations of the first subsystem (described by
canonical coordinates~$x_1$, $y_1$).}

In the following, the Poincar\'e set~$\Bbb P$~\cite{7} plays an essential role. Let
us decompose a disturbing function~$H_1$ in a multiple Fourier series:
$$
H_1 = \sum H^{(m)}(I_1, \ldots , I_n)
\exp [i(m_1 \varphi_1 + \ldots + m_n \varphi_n)]\,, \quad
m = (m_1, \ldots , m_n) \in \Bbb Z^n.
$$
Let~$\omega_i = \frac{dh_i}{dI_i}$ ($1 \le i \le n$) be the frequencies of a
nonperturbed problem; and assume~$\omega = (\omega_1, \ldots , \omega_n)$.
By definition the set~$\Bbb P$ consists of such points~$I = (I_1, \ldots , I_n)
\in \Bbb R^n$,  for which there will be~$n-1$ linearly independent integer
vectors~$\alpha , \alpha ', \ldots \in \Bbb Z^n$, such that

1) $(\omega, \alpha ) = (\omega, \alpha ') = \ldots = 0$;

2) $H^{(\alpha )} (I) \ne 0, \; H^{(\alpha ')} (I) \ne 0, \ldots$.

In a typical situation the set~$\Bbb P$ fills in the range of  values~ $I \in
\Bbb R^n$ \cite{7} everywhere densely.

Our basic result is the following:

{\bf Theorem 1.} {\it Let us assume, that systems with the Hamiltonians~ $h_1, \ldots , h_n$
are non-degenera\-te~$\frac{d^2 h_i}{d I_i^2} \ne 0$, the Poincar\'e set is everywhere dense and
the condition~\thetag{2.7}} {\it is fulfilled. Then}
$$
\rho = c e^{-\beta H_0} (1 + O(\varepsilon ))\,,
\eqno{(2.8)}
$$
{\it where~$c > 0$, $\beta$ are some constants.}

The constant~$c$ is inessential:
the  result of averaging with respect to the measure~\thetag{2.8}
\vspace*{-3mm}
$$
\langle f\rangle  \, = \frac{\int f \rho \, dxdy }{ \int \rho \, dx\,dy }
$$
\vspace*{-5mm}\\
does not depend on this constant. If~$\varepsilon = 0$ the formula~\thetag{2.8} gives the
 Gibbs canonical distribution~\thetag{1.1}.

{\bf Remark 2.} {\it We  stress the fact, that~\thetag{2.8}} {\it is valid
for the fixed  number
of degrees of freedom~$n\ge 2$. A.\,A.\,Vlasov~\cite{8} developed an approach to
the derivation of Gibbs distribution, which would not use the analysis of interaction
of subsystems at all and which is formally suitable for the case~$n = 1$.
This approach is based on the principle of ``maximum statistical independence'',
which seems to be an artificial supposition.}

\section{ Derivation of Gibbs distribution}

The proof of Theorem~1 is based on application of the Poincar\'e method~\cite{9}
in the form indicated in~\cite{7}.

Setting~$\varepsilon = 0$ according to~\thetag{2.6} we obtain that~$\rho_0$ is the
first integral of a complete integrable Hamiltonian system with a Hamiltonian
\vspace*{-3mm}
$$
H_0 = \sum_{k=1}^{n} h_k(I_k)\,.
$$
\vspace*{-5mm}\\
Since nonperturbed system is non-degenerate, i.\,e.
\vspace*{-1mm}
$$
\det \left\| \frac{\partial^2 H_0 }{ \partial I^2} \right\| =
\prod_{k=1}^{n} \frac{d^2 h_k }{ dI_k^2} \ne 0\,,
$$
\vspace*{-3mm}\\
then the function~$\rho_0$
expressed in action-angle variables~$I$, $\varphi \, \text{mod } 2\pi$
 depends only on~$I_1, \ldots , I_n$ \cite{7}.

Furthermore, since~$\rho$ is the first integral of the canonical system of
differential equations with a Hamiltonian~$H$,  their Poisson bracket is equal
to zero:~$\{ \rho , H \} = 0$. Let us differentiate this equality with respect
to~$\varepsilon$ and, then, let us assume~$\varepsilon = 0$. Since~$\rho$ and~$H$
are supposed to be the functions of a class~$C^2$ the differentiations with
respect to phase variables and to the parameter~$\varepsilon$ are commutative.
As a result we obtain the equality
\vspace*{-2mm}
$$
\{ \rho_0 , H_1 \} + \{ \rho_1 , H_0 \} = 0\,,
$$
from which, with the help of the Fourier method and by a known
method~\cite{7}, we can deduce that the functions~$\rho_0(I_1, \ldots , I_n)$ and~$H_o = \sum h_k(I_k)$
are dependent in all points of the Poincar\'e set~$\Bbb P$. According to the
supposition, this set is everywhere dense in a range of action variables~$I$.
Hence, the functions~$\rho_0$ and~$H_0$ are everywhere
dependent by virtue of  continuity.

 The variables~$I_i$ may be presented as single-valued functions
of~$h_i$ in accordance with the assumption made in Section~2 about properties of action-angle
variables. Then~$\rho_0 = \rho_0(h_1, \ldots , h_n)$ and~$H_0 = \sum h_i$. Since
these functions are dependent, $\rho_0$ is a smooth function of~$H_0$.

Indeed, the condition of dependence of~$\rho_0$ and~$H_0$ gives relations
\vspace*{-2mm}
$$
\frac{\partial \rho_0 }{\partial h_i} = \frac{\partial \rho_0 }{ \partial h_j}\,,
\eqno{(3.1)}
$$
\vspace*{-4mm}\\
which are valid for all values~$i,\,j$. If we substitute~$h_n= H_0 - h_1 - \ldots - h_{n-1}$
in the expression for~$\rho_0$, we shall obtain
\vspace*{-2mm}
$$
\rho_0 = f(H_0, h_1, \ldots , h_{n-1}) =
\rho_0\left(h_1, \ldots , h_{n-1}, H_0 - \sum_{k=1}^{n-1} h_k\right)\,.
$$
\vspace*{-3mm}\\
However, this function does not actually depend on~$h_1, \ldots , h_{n-1}$, since
$$
\frac{\partial f }{ \partial h_i} = \frac{\partial \rho_0 }{ \partial h_i} -
\frac{\partial \rho_0 }{ \partial h_n} = 0\,, \quad i < n
$$
\vspace*{-4mm}\\
according to~\thetag{3.1}.
\goodbreak

The probability densities~$p_1, \ldots , p_n$ are the integrals of one-dimensional
systems with Hamiltonians~$h_1, \ldots , h_n$. Hence, in the action-angle
variables, $p_i$ are smooth functions only of~$I_i$. But in this case~$p_i$
is a single-valued differentiable function of the energy~$h_i$.

Thus, the equality~\thetag{2.7} can be presented in the following form~:
$$
\rho_0 (h_1 + \ldots + h_n) = p_1(h_1) \ldots p_n(h_n)\,.
$$
Consecutively differentiating this relation with respect to~$h_1, \ldots , h_n$
and using the positiveness of functions~$p_1, \ldots , p_n$, we obtain equalities
$$
\frac{p_1'}{p_1} = \ldots = \frac{p_n'}{p_n} = -\beta\,,
$$
where~$\beta$ is some constant. Hence,
$$
p_i = c_i e^{-\beta h_i}\,, \qquad c_i = \text{const} > 0\,,
\eqno{(3.2)}
$$
and consequently
$$
\rho_0 = c e^{-\beta H_0}\,, \qquad c = c_1 \ldots c_n > 0\,.
$$
The theorem is proved.

{\bf Remark 3.} {\it As we see from~\thetag{3.2}}, {\it the constant~$\beta$ is one and the same
for all subsystems. It means that the weakly interacting subsystems are in the
thermodynamic equilibrium (and in the limit as~$\varepsilon \to 0$), since
their temperatures~$T = \frac{1}{\beta k}$ are identical.
Thus, when~$\varepsilon \to 0$, the statistical independence of subsystems is
equivalent to their thermodynamic equilibrium. This is the physical sense of
the Gibbs hypothesis.}

\section{Analytical case} In the applications, the Hamiltonian~\thetag{3.2} is
an analytical function of phase variables and the parameter~$\varepsilon$.
In this case it is also natural to consider the probability density~\thetag{2.6}
as an analytical function with respect to~$x$, $y$, $\varepsilon$.

The condition of everywhere-density of the Poincar\'e set can be weakened: it is
enough to require, that~$\Bbb P$ would be a key set for a class of analytical
functions. It means the following: if the analytical function~$f(I)$ is equal to
zero in the points of~$\Bbb P$, then~$f \equiv 0$. The examples of nondense key
sets may be found in~\cite{7}. In particular, if the analytical functions are
dependent in the points~$\Bbb P$, they are everywhere dependent.

{\bf Theorem 2.} {\it
Let us assume, that the systems with Hamiltonians~$h_1, \ldots , h_n$ are
non-degenerate, the Poincar\'e set is a key set for the class of analytical
functions and the equality~\thetag{2.7}} {\it is performed. Then the analytical
density of probability distribution for the system with the
Hamiltonian~\thetag{2.3}}
{\it has the form}
$$
\rho = c e^{-\beta H} [1 + \varepsilon g(H, \varepsilon )]\,,
\eqno{(4.1)}
$$
{\it where~$c > 0$, $\beta \ne 0$ are some constants, $g$ is an analytical
function of the energy~$H$ and the parameter~$\varepsilon$}.

The function $g$ can be represented as a power series in terms of
$\varepsilon$
with coefficients depending on the energy $H$ only.
The summand $\varepsilon g$ in~\thetag{4.1} is completely analogous to
the Gram- 'harlier series in the theory of  distribution  of random variables
which are unessentially different from the normal distributed ones
(see, for example, \cite{10}).

{\it Proof.}
Let us prove Theorem~2. Expand the probability density in
the power series  of $\varepsilon$:
$$
\rho = \rho_0 + \varepsilon \rho_1 + \varepsilon^2 \rho_2 + \ldots \, .
$$
On account of non-degeneracy of the unperturbed system, $\rho_0$ is an
analytical function of action variables $I$ only. Since $\rho_0$ and $H_0$
are  dependent at points $I \in \Bbb P$, and $\Bbb P$ is the key set,
$\rho_0$ and $H_0$ are dependent everywhere.
Therefore, taking into consideration the Gibbs hypothesis \thetag{2.7},
$$
\rho_0 = c e^{-\beta H_0}\,; \quad c, \beta = \text{const},
$$
(see Section~3). Since $\rho_0 > 0$, we have $c > 0$. The constant
$\beta$ is not equal to zero, otherwise
 $\rho_0$ is not the  density of probabilistic measure
 (the volume of the whole phase space is infinite).

Further, the analytical function
$$
\frac{\rho }{ ce^{-\beta H}  }
$$
is an integral of Hamilton equations with the Hamiltonian \thetag{2.7}.
Let us expand this function in power series of $\varepsilon$:
$1 + \varepsilon G_0 + \varepsilon ^2 G_1 + \ldots$ . It is obvious that
 the series
$$
G_0 + \varepsilon G_1 + \ldots
\eqno{(4.2)}
$$
is an integral of the same system. With the help of the method used in
 Section~3 one can prove that~$G_0$ is an analytical function of $H_0$:
 $G_0 = g_0(H_0)$.
It is obvious that  the power series
$$
\frac{[G_0 + \varepsilon G_1 + \ldots - g_0(H)]}{\varepsilon} =
F_0 + \varepsilon F_1 + \ldots
$$
is again a first integral. Thus, $F_0$ is an analytical function of $H_0$:
$F_0 = g_1(H_0)$. Extending infinitely this procedure, we obtain
that  the series \thetag{4.2}  has actually the form:
$$
g_0(H) + \varepsilon g_1(H) + \ldots \, .
$$
Denoting this function by $g(H, \varepsilon)$, we obtain the desired
formula \thetag{4.1}. $\blacksquare$

{\bf Remark 4.} {\it Let us introduce  $K(H, \varepsilon)$ by setting}
$$
\rho = c e^{-\beta K},
\eqno{(4.3)}
$$
{\it where the density $\rho$ is given by~\thetag{4.1}}. {\it The function $K$
can be expanded in power series $K_0 + \varepsilon K_1 + \ldots$, where}
$$
K_0 = H_0\,, \quad K_1 =H_1 - \frac{g_0(H_0)}{\beta}, \ldots \, .
$$
{\it Hamilton equations with the Hamiltonians $H$ and $K$ possess the same
trajectories, however, times of motion along these trajectories are different.
The formula \thetag{4.3}} {\it implies the Gibbs canonical distribution for
a slightly changed Hamiltonian system.}

\section{Application to the system of weakly connected pendula}
Let us consider $n$ identical mathematical pendula of  mass $m$ and
 length $l$,   consecutively connected with each other by elastic springs
with small rigidity $\varkappa$. For the simplicity we assume that
the fixed points of pendula coincide.  This system with
 $n$ degrees of freedom is described by canonical differential equations
with the Hamiltonian
$H_0 + \varepsilon H_1$, where
$$
H_0 = \sum_{i=1}^{n} \frac{y_i^2 }{2ml^2} - mgl\cos x_i\,, \qquad
H_1 = \sum_{i=1}^{n-1} \cos (x_i - x_{i+1})\,,
$$
$\varepsilon = -\frac{\varkappa l^2}{4}$ is a small parameter.
The transition to angle-action variables is carried out for the pendulum
with the help of
elliptic functions (see, for example, \cite{7}).
It is possible to show that the Fourier series of perturbation function with respect
to angle variables $\varphi_1, \ldots , \varphi_n \, \text{mod } 2\pi$
has the form:
$$
\begin{aligned}
H_1 = \sum_{m_1, m_2} h_{m_1, m_2} (I_1) e^{2i(m_1 \varphi_1 + m_2 \varphi_2)} +
\ldots
+ \sum_{m_{n-1}, m_n} h_{m_{n-1}, m_n} (I_n)
e^{2i(m_{n-1} \varphi_{n-1} + m_n \varphi_n)}\,.
\end{aligned}
$$
The summation is taken over all integer  $m_1, \ldots , m_n$ from
$-\infty$ to $+\infty$. The coefficients in this expansion can be expressed
explicitly with the help of known expansions of elliptic functions
$s n^2$, $c n^2$ and $sn cn$ in
Fourier series \cite{11}. All of them are not equal to zero.

The Poincar\'e set $\Bbb P$ is defined in this problem  as   the set of points
$I = (I_1, \ldots , I_n)$, satisfying  $n-1$  equations
$$
m_1 \omega_1(I_1) + m_2 \omega_2(I_2) = \ldots =
m_{n-1} \omega_{n-1}(I_{n-1}) + m_{n} \omega_{n}(I_{n}) = 0\,,
$$
where  either $m_1 m_2 \ldots m_{n-1} \ne 0$ or $m_2 m_3 \ldots m_{n} \ne 0$.
If one of the two latter  conditions is fulfilled,
the vectors $(m_1, m_2, 0, \ldots , 0)$, $(0, m_2, m_3, 0, \ldots , 0)$, \ldots ,
$(0, 0, \ldots, m_{n-1}, m_n)$ are linearly independent.
It is possible to show that the set  $\Bbb P$, which consists of infinitely many
curves,    fills in the domain of definition of action variables
 $\{ I_1 \ge 0, \ldots , I_n \ge 0 \}$ everywhere densely.

Thus, the statements of  Theorems~1 and 2 are valid for the chain of
connected pendula. If one of the springs is taken away, the
system splits into two disconnected chains, the Hamiltonians of which are
the first integrals of the total system. In this case the
density  of probability  distribution
is not the function of the total energy of the system only and thus is not
subjected to the Gibbs distribution.

\section{Thermodynamics of mechanical systems}
Let us consider  a Hamiltonian system with the Hamiltonian $H$, where
the density  of probability  distribution is defined by the Gibbs formula
\thetag{1.1}.  According to theorem~1, it can be a system with $n$
degrees of freedom, composed of independent one-dimensional subsystems,
or one of these subsystems. We shall show, after Gibbs \cite{1}, that
 this   mechanical system   can be naturally connected with
some thermodynamic system.

Let us assume that the Hamiltonian $H$ depends not only on
canonical variables $x$, $y$, but also on several parameters
$\lambda_1, \ldots , \lambda_m$.  One can take, for example,
the length of the pendulum as such a parameter.
The parameters  $\lambda$ can be considered as generalized coordinates
of some system with $n+m$ degrees of freedom and  the constancy of
$\lambda$ as  application of holonomic constraints. Thus, let
$H_{*} (x, y, \lambda , \mu )$ be  the Hamiltonian of a system with
 $n+m$ degrees of freedom, $\mu_1, \ldots , \mu_m$ be canonical momenta
conjugated with additional coordinates $\lambda_1, \ldots , \lambda_m$.
The dynamics of the extended system is described by the canonical equations
$$
\dot x = \frac{\partial H_{*} }{ \partial y}\,, \qquad
\dot y = -\frac{ \partial H_{*} }{ \partial x}\,, \qquad
\dot \lambda = \frac{\partial H_{*} }{ \partial \mu} \,, \qquad
\dot \mu = - \frac{\partial H_{*} }{ \partial \lambda} \,.
\eqno{(6.1)}
$$

Let us impose  $m$ independent  relations
$\lambda_1, \ldots , \lambda_m = \text{const}$ on this system.
Therefore, $\dot \lambda_i = 0$, and from the equations
$$
\frac{\partial H_{*} }{ \partial \mu_1} = \ldots =
\frac{\partial H_{*} }{ \partial \mu_m} = 0
\eqno{(6.2)}
$$
one can derive momenta  $\mu$ as functions of $x$, $y$ and constant parameters
$\lambda$. The sufficient condition of solvability of the equations  \thetag{6.2}
with respect to $\mu$ is reduced to the inequality
$$
\det \left\| \frac{\partial^2 H_{*} }{ \partial \mu_i \partial \mu_j} \right\| \ne 0\,.
$$
It is fulfilled automatically  if $H_{*}$ is a positively defined
quadratic form of $n+m$ momenta $y$, $\mu$.
\goodbreak

Substituting the obtained expressions for $\mu$ in the last
set of equations \thetag{6.1}, we obtain  additional relations on
canonical variables $x$, $y$, which, of course,  are not fulfilled in
general case. Therefore, it is necessary to introduce additional forces,
the constraint reactions  $R_1, \ldots , R_m$, and replace the latter equation
of  \thetag{6.1} by
$$
\dot \mu = - \frac{\partial H_{*} }{ \partial \lambda} + R\,.
\eqno{(6.3)}
$$

Let $\mu = \mu(x, y, \lambda )$ be a solution of the algebraic system
\thetag{6.2}. We assume that
$$
H(x, y, \lambda ) = H_{*} (x, y, \lambda , \mu (x, y, \lambda )).
$$
On account of \thetag{6.2},
$$
\frac{\partial H }{ \partial x} = \frac{\partial H_{*} }{ \partial x}\,, \qquad
\frac{\partial H }{ \partial y} = \frac{\partial H_{*} }{ \partial y}\,, \qquad
\frac{\partial H }{ \partial \lambda} = \frac{\partial H_{*} }{ \partial \lambda}\, .
$$
Consequently, for constant values of $\lambda$ the variables $x$, $y$
change  according to Hamilton equations with the Hamiltonian
$H$, and the equation \thetag{6.3} can be replaced by
$$
\dot \mu = - \frac{\partial H }{ \partial \lambda} + R\,.
\eqno{(6.4)}
$$

Let $x(t, x_0, y_0)$, $y(t, x_0, y_0)$ be solutions of canonical equations
with the Hamiltonian $H$. We assume that for every such solution
$\mu (t) / t \to 0$ as $t \to +\infty$. Then the time-average of
$\dot \mu$ is equal to zero. This assumption is automatically fulfilled if
the configuration space $\{ x \}$ is compact:
on account of existence of energy integral the function $\mu (t)$ is bounded.

At first let us average  both parts of the equality \thetag{6.4} with respect to
time and then with respect to the measure $\rho (x_0, y_0) dx_0 dy_0$,
where $\rho$ is given by \thetag{1.1}. According to the Birkhoff-Khinchin ergodic
theorem \cite{12} the obtained relation is equivalent to the following one:
$$
\langle R\rangle  =\left\langle  \frac{\partial H }{ \partial \lambda} \right\rangle \,.
\eqno{(6.5)}
$$
Here $\langle  \; \rangle $ is the mean with respect to the Gibbs measure \thetag{1.1}.

{\bf Remark 5.} {\it
Usually \cite{3, 4} the relation \thetag{6.5}} {\it is derived from the simplified
relation of \thetag{6.4}}, {\it which lacks the derivative}  $\dot \mu$.

The relation \thetag{1.2} defines the ``internal'' energy $E$ as a function
of parameters  $\lambda_1, \ldots , \lambda_m$ and
$\beta$ (recall that $\beta^{-1} = kT$). Let us set
$\Lambda = -\langle R\rangle ,$
i.\,e., the phase mean values of the above introduced constraint reactions
(with the opposite sign). The relations
$$
\Lambda_i = f_i(\lambda_1, \ldots , \lambda_m, \beta )\,, \quad
1 \le i \le m
\eqno{(6.6)}
$$
are usually called in thermodynamics the  equations of state. Assignment of
functions $E$ and $\Lambda_i$ is included in the definition of thermodynamic system.
However, these functions can not be arbitrary, since the first and the second
principles of thermodynamics must be fulfilled.

Let us introduce the statistical integral
$$
Z(\lambda , \beta ) = \int e^{-\beta H}\, dx\, dy\,.
$$
One can verify the validity of the equalities \cite{3, 4}
$$
\Lambda_i = \frac{1}{\beta} \frac{\partial \ln Z}{\partial \lambda_i}, \quad
(1 \le i \le m)\,, \qquad
E = - \frac{\partial \ln Z}{\partial \beta}\,.
\eqno{(6.7)}
$$

In thermodynamics the main role plays the differential 1-form of heat gain
$$
\omega = dE + \sum \Lambda_i \,d\lambda_i\,.
$$
Motivations for this definition can be found in \cite{2, 4}. According to
the first equality of \thetag{6.7}
the form $\omega$ is the exact differential for fixed values of $\beta$
(or absolute temperature $T$).
This is the first principle of thermodynamics. Then, taking into account~\thetag{6.7}
we obtain
$$
\beta \omega = \beta\, dE + \sum \Lambda_i\, d\lambda_i =
d(\beta E) - E \,d\beta + \sum \Lambda_i\, d\lambda_i =
d(\beta E + \ln Z)\,.
$$
Thus, the form of heat gain possesses the integrating factor
$\beta = 1/(kT)$. This is the second principle of thermodynamics. The function
$S = \beta E + \ln Z$ is the entropy of thermodynamic system.

Finally, let us give an illustrative example. We consider a  motion of
a point of mass $m$, attached to the end of unstretchable thread of length
 $l$. Let $x$ be the  angle of rotation  of the thread, $y$ be the conjugate
canonical momentum. If active forces are not applied to the point,
its dynamics is described by canonical equations with the Hamiltonian
$$
H = \frac{y^2 }{ 2 m l^2}\,.
$$
Let us take the length of the thread $l$ as the parameter $\lambda$.
The statistical integral is equal to
$$
Z = (2 \pi )^{3/2} m^{1/2} l \beta^{-1/2}.
$$
Let $p$ be a force, corresponding to the parameter $l$ (the thread tension).
Then, according to \thetag{6.7} the  equation of state \thetag{6.6}
has the form $p = 1/(\beta l)$ or $pl = kT$. It is similar to the
ideal gas law. The second equation \thetag{6.7} gives the
relation for the internal energy
 $E = 1/(2\beta )$ or $E = kT/2$. From these relations we obtain the formula
 $p = 2E/l$, which, however, was  valid before the application of
averaging procedure.

The work is partly supported by RFBR (No.~99--01--01096) and
INTAS (No.~96--0793).

\end{document}